\documentclass[preprint,prc,aps,showpacs,superscriptaddress]{revtex4}
\usepackage{epsfig}
\usepackage{amsmath}
\usepackage[hyperref]{hyperref}

\newcommand{\Pomeron}{I\!\!P}

\begin{document}

\title{Color fluctuation approximation for multiple interactions in 
leading twist theory of nuclear shadowing}

\author{V.~Guzey}
\email{vguzey@jlab.org}
\affiliation{Theory Center, Thomas Jefferson National Accelerator Facility, 
Newport News, Virginia 23606, USA}

\author{M.~Strikman}
\email{strikman@phys.psu.edu}
\affiliation{Department of Physics, Pennsylvania State University,
University Park, Pennsylvania 16802, USA}

\preprint{JLAB-THY-09-1048}
\pacs{24.85.+p}

\begin{abstract}

The leading twist theory of nuclear shadowing predicts the shadowing correction to nuclear parton distributions at small $x$ by connecting it to the leading twist 
hard diffraction in electron-nucleon scattering.
The uncertainties of the predictions are related to the shadowing effects 
resulting from the interaction of the hard probe 
with $N \ge 3$ nucleons. We argue that the pattern of hard diffraction observed at HERA 
allows one to reduce these uncertainties.
We develop a new approach to the treatment of these multiple 
interactions, which is based on the concept of the color fluctuations 
and accounts for the presence 
of both point-like and hadron-like configurations in the virtual photon wave function.
Using the developed framework, we update our 
predictions for  
the leading twist nuclear shadowing in nuclear parton distributions of heavy nuclei 
at small $x$.

\end{abstract}

\maketitle

\section{Introduction}
\label{sec:intro}

In this work, we consider the quark and gluon parton distribution functions (PDFs) in 
nuclei at small values of Bjorken $x$ and their
reduction as compared
with the incoherent sum of the nucleon PDFs because of the phenomenon
of nuclear shadowing.
Most of experimental information on nuclear PDFs comes from inclusive 
deep inelastic scattering (DIS)
with nuclear targets which measures the nuclear structure function $F_{2A}(x,Q^2)$.
For $x < 0.05$, $F_{2A}(x,Q^2)< A F_{2N}(x,Q^2)$, which is called
nuclear shadowing [$F_{2N}(x,Q^2)$ is the 
isoscalar nucleon structure function and $A$ is the number of nucleons].
Because of the factorization theorem for DIS
(for a review, see Ref.~\cite{Brock:1993sz}), which relates 
$F_{2A}(x,Q^2)$ to nuclear parton distributions $f_{j/A}(x,Q^2)$
($j$ is the parton flavor), nuclear shadowing is also present in nuclear PDFs,
$f_{j/A}(x,Q^2) < Af_{j/N}(x,Q^2)$ for $x < 0.05$, where $f_{j/N}(x,Q^2)$ is the PDF of the
free nucleon. This finds evidence in the results of the 
global fits that extract nuclear PDFs 
from various  data on hard scattering with nuclei~\cite{Eskola:1998iy,Eskola:1998df,Eskola:2002us,Eskola:2008ca,Eskola:2009uj,Hirai:2001np,Hirai:2004wq,Hirai:2007sx,deFlorian:2003qf}.

Nuclear PDFs at small $x$ 
play an important role in the phenomenology of hard scattering
with nuclei. 
Their knowledge 
is required for the evaluation and interpretation of hard phenomena in proton-nucleus and nucleus-nucleus collisions at Relativistic Heavy Ion Collider (RHIC) and the Large Hadron Collider (LHC),
in real photon-nucleus interactions in ultraperipheral collisions at the LHC~\cite{Baltz:2007kq}, 
and in lepton-nucleus scattering at the future Electron-Ion Collider (EIC)~\cite{Deshpande:2005wd,eic}.
In addition, nuclear PDFs at small $x$ are needed for the quantitative estimation 
of the onset of saturation in ultra high energy
interactions with nuclei, which can be studied at the LHC and the EIC.

A comparison of the results of
the  global fits for nuclear PDFs obtained
 by various groups~\cite{Eskola:1998iy,Eskola:1998df,Eskola:2002us,Eskola:2008ca,Eskola:2009uj,Hirai:2001np,Hirai:2004wq,Hirai:2007sx,deFlorian:2003qf} shows significant discrepancies in the
predictions for nuclear PDFs at small $x$
(uncertainties of individuals fits at small $x$ are also very 
large~\cite{Eskola:2009uj,Hirai:2007sx}).
 The main reason for this is that the global
fits are predominantly based on fixed-target data that do not cover
the small-$x$ region (by requiring
 that $Q^2 > 1$ GeV$^2$ is sufficient for the  applicability 
of the factorization theorem, one limits $x > 5 \times 10^{-3}$).
In addition, the gluon nuclear PDF is determined indirectly from the scaling violations
using the very limited data.
Therefore, the extrapolation of the obtained
nuclear PDFs 
to the low values of Bjorken $x$ that will be probed at the LHC and the
EIC is essentially uncontrolled. An alternative to the global fits is provided by 
the approaches that
attempt to predict nuclear shadowing for nuclear PDFs 
using the high-energy dynamics of the strong interactions.

We use the so-called leading twist theory of nuclear shadowing~\cite{Frankfurt:1998ym}. 
It combines the technique used by Gribov 
to derive nuclear shadowing 
for the total hadron-deuteron cross section 
at high energies~\cite{Gribov:1968jf} and 
the QCD factorization theorems for inclusive~\cite{Brock:1993sz} and diffractive~\cite{Collins:1997sr} DIS.
The numerical predictions employ the results of the leading twist 
QCD analyses of hard diffraction 
in lepton-proton DIS at HERA~\cite{Aktas:2006hy,Aktas:2006hx,Chekanov:2008cw}.
Although in the leading twist approach
the hard probe interacts with one parton of the nucleus,
in the target rest frame, nuclear shadowing appears as the effect of multiple
interactions of the projectile (virtual photon) with 
several (all) nucleons of the target. 
The interaction with $N=2$ nucleons is related in a model-independent way to the 
diffractive PDFs of the nucleon.
The account of the interaction with  $N \geq 3$
nucleons is model-dependent and sensitive to the underlying 
dynamics of the hard diffraction. 
The recent HERA data~\cite{Aktas:2006hy,Aktas:2006hx,Chekanov:2008cw} 
revealed that the energy dependence of the hard diffraction in DIS 
(dependence on the light-cone fraction $x_{\Pomeron}$) 
is close to that of the soft processes. This 
indicates that the hard diffraction in DIS is dominated by 
large-size hadron-like configurations in the photon wave function.
This observation allows us to improve the treatment of the contribution to nuclear shadowing 
coming from the interactions with $N\ge 3$ nucleons as compared to the simplified quasi-eikonal approximation used in our earlier papers~\cite{Frankfurt:2003zd}
by taking into account the presence 
of both point-like and hadron-like configurations in the virtual photon.
The goal of the present Letter is to present a 
new, improved treatment of 
such multiple interactions
and to update predictions for nuclear shadowing in nuclear PDFs.

We emphasize that the presence of the small-size (point-like) configurations and, in general,
configurations of different transverse sizes that interact with different cross sections
(we call such configurations color or cross section fluctuations)
is much more important for the virtual photon
than for hadronic projectiles.

Our Letter is organized as follows. In Sec.~\ref{sec:lt_theory}, we 
briefly review 
the leading twist theory of nuclear shadowing and present a new formalism 
for the treatment of the multiple interactions of the virtual photon
 with the nucleons of the nuclear target, which
is based on the concept of color (cross section) fluctuations.
In Sec.~\ref{sec:cfa}, we show that the complete treatment of color fluctuations can be well approximated by the so-called color
fluctuation approximation. In Sec.~\ref{sec:predictions}, we present 
updated predictions 
for the effect of nuclear shadowing in nuclear parton distributions. 
We focus on the predictions for  heavy nuclei
at small $x$ since modifications of the predictions for light nuclei
(where the $N=2$ term dominates) 
are rather small.
Our results are summarized in Sec.~\ref{sec:summary}.

\section{Leading twist theory of nuclear shadowing and cross section fluctuations for multiple interactions}
\label{sec:lt_theory}

The phenomenon of nuclear shadowing is fairly well-understood: in the target rest
frame, nuclear shadowing arises as the result of multiple interactions of the 
projectile (virtual photon) with several nucleons of the nuclear target.
The number of the interactions increases with decreasing Bjorken $x$, which is
a result of the space-picture of the strong interactions, see e.g., 
Ref.~\cite{Gribov:1973jg}. At sufficiently high energies (small Bjorken $x$),
the virtual photon can interact with all 
the nucleons of the target that are located 
in the photon's path.

\begin{figure}[t]
\begin{center}
\epsfig{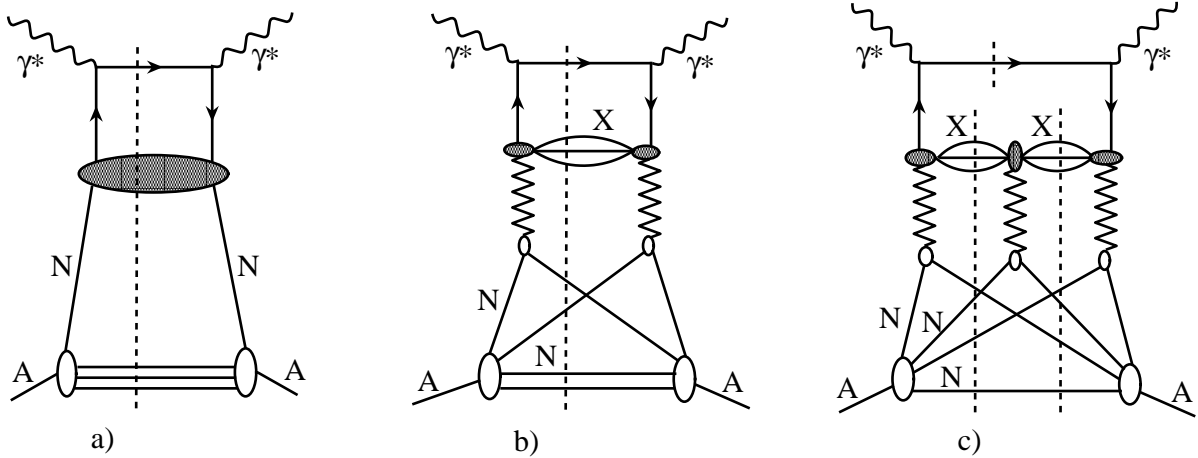}
\caption{Graphical representation of the nuclear structure function $F_{2A}(x,Q^2)$. 
Graphs $a$, $b$ and $c$ correspond to the interaction with one, two and three nucleons, 
respectively.
The latter two graphs and the interaction with four and more nucleons (not shown) lead
to nuclear shadowing.
The dashed vertical lines represent taking of the 
imaginary part.
}
\label{fig:Master1}
\end{center}
\end{figure}

The graphs that contribute to the nuclear structure function $F_{2A}(x,Q^2)$
are presented in Fig.~\ref{fig:Master1}, where we used the optical theorem to 
relate the imaginary part of the $\gamma^{\ast} A$ forward scattering amplitude to the 
nuclear structure function. In the figure, graphs $a$, $b$, and $c$ correspond to 
the interaction
with one, two, and three nucleons of the nuclear target, respectively.
Note that the graphs for the interaction with four or more nucleons are not shown, but
assumed. Graphs $b$, $c$ and higher scattering terms are responsible for  nuclear shadowing
in $F_{2A}(x,Q^2)$.

The contribution of graph $a$, which is conveniently denoted
 $F_{2A}^{(a)}(x,Q^2)$, is 
\begin{equation}
F_{2A}^{(a)}(x,Q^2)=A F_{2N}(x,Q^2) \,,
\label{eq:m2}
\end{equation}
where $F_{2N}(x,Q^2)$ is the isospin-averaged structure function of the nucleon. 
In Eq.~(\ref{eq:m2}), we neglected the deviation from the many-nucleon approximation for the description of nuclei and the Fermi motion effect, which are numerically unimportant at small $x$.

The calculation of the contribution of graph $b$, $F_{2A}^{(b)}(x,Q^2)$,
is fairly straightforward, but lengthy~\cite{Gribov:1968gs}.
The detailed derivation, including the effect of the real part of the 
diffractive amplitude, is given in
Ref.~\cite{Frankfurt:2003zd}. Here we present 
the final result for $F_{2A}^{(b)}(x,Q^2)$,
\begin{align}
F_{2A}^{(b)}(x,Q^2)&=
-8 \pi A(A-1) \Re e \frac{(1-i\eta)^2}{1+\eta^2} B_{\rm diff}\int^{0.1}_x d x_{\Pomeron}
F_2^{D(3)}(x,Q^2,x_{\Pomeron}) \nonumber\\
&\times \int d^2 \vec{b}  
\int^{\infty}_{-\infty}d z_1 \int^{\infty}_{z_1}d z_2 \rho_A(\vec{b},z_1) \rho_A(\vec{b},z_2) 
e^{i (z_1-z_2) x_{\Pomeron} m_N}   \,,
\label{eq:m11}
\end{align}
where $F_2^{D(3)}$ is the nucleon diffractive structure function measured in hard $\gamma^{\ast} p$
inclusive diffraction; $\rho_A$ is the nuclear density; $\eta \approx 0.17$ is the ratio of the
real to imaginary parts of the  $\gamma^{\ast} p$ diffractive amplitude;
$B_{\rm diff}=6$ GeV$^{-2}$ is the slope of the $t$ dependence of the diffractive $\gamma^{\ast}p$ cross section;
$x_{\Pomeron}$ is the light-cone fraction of the nucleon momentum
 carried by the Pomeron (see the discussion below); 
$m_N$ is the nucleon mass.

The nuclear density $\rho_A$ depends on the transverse coordinate (impact parameter), $\vec{b}$,
and the longitudinal coordinates, $z_1$ and $z_2$, of the interacting nucleons.
The ordering $z_2 > z_1$ follows from the space-time evolution of the scattering process.
The $e^{i (z_1-z_2) x_{\Pomeron} m_N}$ factor accounts for the excitation of the
intermediate diffractive state denoted by $X$ in Fig.~\ref{fig:Master1}. 
At high energies, 
the $\gamma^{\ast}N$ interaction that leads to nuclear shadowing is diffractive 
in character. This is represented by the zigzag lines in Fig.~\ref{fig:Master1}. 
It is convenient to think of the zigzag lines as 
depicting  effective Pomeron exchanges. 
In this case, $x_{\Pomeron}$ represents
the light-cone fraction of the nucleon momentum
 carried by the Pomeron, $x_{\Pomeron} \approx (M_X^2+Q^2)/(W^2+Q^2)$, where 
$M_X$ is the invariant mass of the diffractive state $X$, and 
$W$ is the invariant $\gamma^{\ast}p$ energy.
The lower limit of integration over $x_{\Pomeron}$ in Eq.~(\ref{eq:m11}) 
corresponds to
$M_X=0$; the upper limit is determined by the typical cut on  $M_X$,
$M_X^2 \leq 0.1 W^2$, which arises because of the nuclear form factor. 
Note that Eq.~(\ref{eq:m11})
 is valid independently of the validity of the 
leading twist approximation for the hard diffraction.

\begin{figure}[t]
\begin{center}
\epsfig{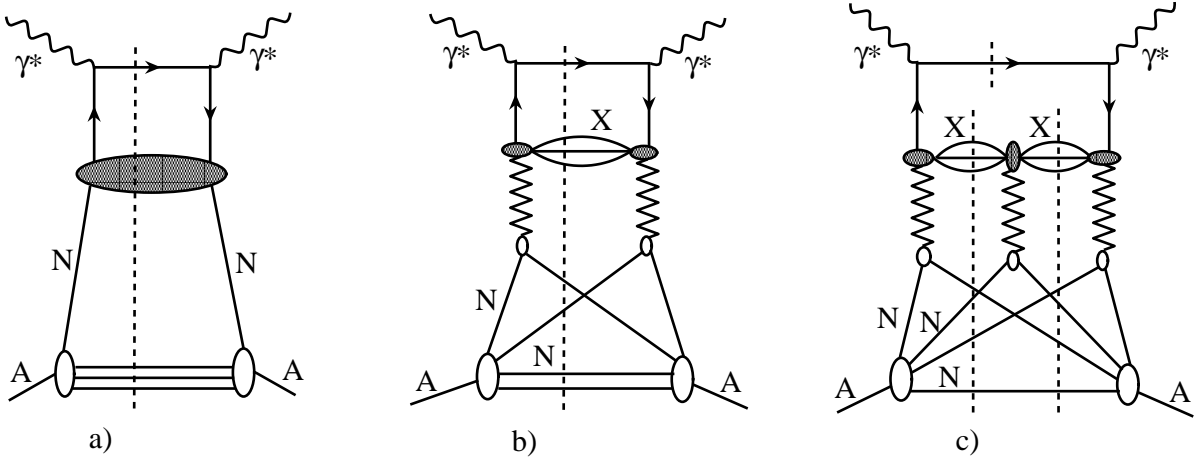}
\caption{Graphical representation of the multiple scattering series for
the quark distribution in a nucleus.
Graphs $a$, $b$ and $c$ correspond to the interaction with one, two and three nucleons, respectively.
The latter two graphs and the interaction with four and more nucleons (not shown) lead
to nuclear shadowing.
}
\label{fig:Master1b_2009}
\end{center}
\end{figure}

One of the key features of the leading twist 
theory  of nuclear shadowing~\cite{Frankfurt:1998ym,Frankfurt:2003zd} is the possibility to predict nuclear
shadowing at the level of parton distributions. Using the QCD factorization theorems for 
inclusive DIS and hard diffraction in DIS,
one can replace the observable structure functions by the corresponding parton distribution.
This is shown in Fig.~\ref{fig:Master1b_2009}, which represents 
the multiple scattering series for the quark 
distribution in nuclei. A similar graphical representation can also be given  
for the gluon distribution 
using a hard probe directly coupled to gluons.

The contribution of graph $a$ in Fig.~\ref{fig:Master1b_2009}, 
which we denote $f_{j/A}^{(a)}(x,Q^2)$, is readily obtained from Eq.~(\ref{eq:m2}),
\begin{equation}
x f_{j/A}^{(a)}(x,Q^2)=A \,x f_{j/N}(x,Q^2) \,.
\label{eq:m2_c}
\end{equation}
The contribution of graph $b$ is obtained from Eq.~(\ref{eq:m11}),
\begin{align}
x f_{j/A}^{(b)}(x,Q^2)&=
-8 \pi A(A-1) \Re e \frac{(1-i\eta)^2}{1+\eta^2} B_{\rm diff}\int^{0.1}_x d x_{\Pomeron}
\beta f_j^{D(3)}(\beta,Q^2,x_{\Pomeron}) \nonumber\\
&\times \int d^2 \vec{b}  
\int^{\infty}_{-\infty}d z_1 \int^{\infty}_{z_1}d z_2 \,\rho_A(\vec{b},z_1) \rho_A(\vec{b},z_2) 
e^{i (z_1-z_2) x_{\Pomeron} m_N}   \,,
\label{eq:m12}
\end{align}
where $f_j^{D(3)}$ is the diffractive parton distribution of flavor $j$ in a proton.
According to the factorization theorem~\cite{Collins:1997sr}, $f_j^{D(3)}$ is a leading-twist
distribution, whose $Q^2$ evolution is given by the DGLAP equations. This is supported by the 
analyses of diffraction at HERA~\cite{Aktas:2006hy,Aktas:2006hx,Chekanov:2008cw}. Therefore, the contribution of $x f_{j/A}^{(b)}(x,Q^2)$
to nuclear shadowing is also a leading twist function,
which gives the name to the present approach---the leading twist 
theory of nuclear shadowing.

The derivation of the expressions for $xf_{j/A}^{(a)}$ and $xf_{j/A}^{(b)}$ is general
and model-independent: the only simplifying approximations are the neglect of
nucleon correlations in the nuclear wave function and of the $t$ dependence
of the elementary diffractive $\gamma^{\ast}N \to X N$ amplitude.

Graph $b$ in Figs.~\ref{fig:Master1} and 
\ref{fig:Master1b_2009} approximates well nuclear shadowing in the low 
nuclear density limit, when the interaction with only two nucleons is important. 
As one decreases $x$, graph $c$ and higher rescattering terms also become progressively 
important. 
To evaluate their contribution, one needs to model 
the interaction of the intermediate state $X$ with the nucleons of the target.
Our approach is based on the high-energy
formalism of cross section 
fluctuations~\cite{Feinberg56,Good:1960ba,Blaettel:1993ah,Frankfurt:2000tya},
which provides a good description of the total hadron-nucleus cross sections 
and, which is far less trivial, of the coherent inelastic diffraction in hadron-nucleus scattering; for a review and references, see Ref.~\cite{Frankfurt:2000tya}.
In this formalism, the wave function of a 
fast projectile (virtual photon) 
is expanded in terms of eigenstates of the scattering operator, $|\sigma \rangle$.
Each eigenstate interacts with target nucleons with a certain cross section $\sigma$.
The probability for the incoming virtual photon to fluctuate in a given eigenstate
is given by the distribution $P_j(\sigma)$. We explicitly show the dependence
of $P_j(\sigma)$ on the parton flavor $j$ as a reminder that DIS
probes a particular parton distribution of the target.

The entire series of multiple interactions shown in Fig.~\ref{fig:Master1b_2009}
can be summed by the standard Glauber formalism generalized to include 
cross section fluctuations, see, e.g., Ref.~\cite{Frankfurt:2000tya}.
Assuming that $A \gg 1$ such that the multiple interactions can be exponentiated, we obtain
\begin{align}
xf_{j/A}&(x,Q^2)=\frac{xf_{j/N}(x,Q^2)}{\langle \sigma \rangle_j}2\, \Re e \int d^2 b \left\langle 
\left(1-e^{-\frac{A}{2}(1-i\eta) \sigma T_A(b)}\right) \right\rangle_j
\nonumber\\
&=A xf_{j/N}(x,Q^2)- \frac{xf_{j/N}(x,Q^2)}{\langle \sigma  \rangle_j} 2 \Re e \int d^2b \frac{\sum_{k=2}^{\infty}(-\frac{A}{2}(1-i\eta) T_A(b))^k \langle \sigma^k \rangle_j}{k!} \,,
\label{eq:fluct1}
\end{align}
where $T_A(b)=\int^{\infty}_{-\infty} dz \rho_A(b,z)$; 
$\langle \dots \rangle_j$ denotes
the integration over $\sigma$ with the weight $P_j(\sigma)$.
The interaction with $k$ nucleons 
probes the $k$th moment of the distribution $P_j(\sigma)$,
$\langle \sigma^k \rangle_j=\int_0^{\infty} d \sigma P_j(\sigma) \sigma^k$.

Equation~(\ref{eq:fluct1}) is valid at small $x$ (high energies), when the effect of 
the finite coherence length (the coherence length is proportional
to the lifetime of the fluctuations $|\sigma \rangle$) is unimportant. In this case, the 
$e^{i(z_1-z_2) m_N x_{\Pomeron}}$ factor in Eq.~(\ref{eq:m12}) 
can be set to unity.

The contribution to nuclear shadowing from the interaction with $N=2$ nucleons is related in a model-independent way to the  diffractive PDFs of the nucleon.
This means that 
$\langle \sigma^2 \rangle_j$ is proportional to the diffractive parton distribution~\cite{Frankfurt:1998ym,Frankfurt:2003zd},
\begin{equation}
\frac{\langle \sigma^2 \rangle_j}{\langle \sigma \rangle_j} \equiv \sigma_{2}^j(x,Q^2)
=\frac{16 \pi B_{\rm diff}}{(1+\eta^2) x f_{j/N}(x,Q^2)}
\int^{0.1}_{x} d x_{\Pomeron} \beta f_j^{D(3)}(\beta,Q^2,x_{\Pomeron}) \, .
\label{eq:m17}
\end{equation}
Therefore, Eq.~(\ref{eq:fluct1}) can be written as
\begin{align}
xf_{j/A}(x,Q^2)&=Axf_{j/N}(x,Q^2) \nonumber\\
-xf_{j/N}(x,Q^2) &\sigma_{2}^j(x,Q^2)2\, \Re e \int d^2 b 
\frac{\left\langle 
\left(e^{-\frac{A}{2}(1-i\eta) \sigma T_A(b)}-1+\frac{A}{2}(1-i\eta) \sigma T_A(b)\right) \right\rangle_j}{\langle \sigma^2 \rangle_j}
\,.
\label{eq:fluct2}
\end{align}
In order to cast Eq.~(\ref{eq:fluct2}) into the more standard form~\cite{Frankfurt:2003zd},
we reintroduce the dependence on the longitudinal coordinates $z_1$ and $z_2$, 
use the definition of $\sigma_{2}^j(x,Q^2)$ from Eq.~(\ref{eq:m17}),
and identically
rewrite Eq.~(\ref{eq:fluct2}) in the following form,
\begin{align}
xf_{j/A}(x,Q^2)&=Axf_{j/N}(x,Q^2) \nonumber\\
&-8 \pi A^2 \Re e \frac{(1-i\eta)^2}{1+\eta^2}
B_{\rm diff}
\int^{0.1}_{x} d x_{\Pomeron} \beta f_j^{D(3)}(\beta,Q^2,x_{\Pomeron})
\nonumber\\
&\times
\int d^2 b  \int^{\infty}_{-\infty}d z_1 \int^{\infty}_{z_1}d z_2 \rho_A(\vec{b},z_1) \rho_A(\vec{b},z_2) 
 \frac{\left\langle \sigma^2 e^{-\frac{A}{2} (1-i\eta) \sigma \int_{z_1}^{z_2} dz^{\prime} \rho_A(\vec{b},z^{\prime})} \right\rangle_j}{\langle \sigma^2 \rangle_j}
\,.
\label{eq:fluct2_b}
\end{align}
Finally, we restore the effect of the finite coherence length by reintroducing the 
$e^{i (z_1-z_2) x_{\Pomeron} m_N}$ factor,
replace $A^2$ by $A(A-1)$ to have the correct number of the nucleon pairs,
and obtain 
our general
expression for the nuclear parton distribution modified by nuclear shadowing,
\begin{align}
&xf_{j/A}(x,Q^2)=Axf_{j/N}(x,Q^2) \nonumber\\
&-8 \pi A (A-1) \Re e \frac{(1-i\eta)^2}{1+\eta^2}
B_{\rm diff}
\int^{0.1}_{x} d x_{\Pomeron} \beta f_j^{D(3)}(\beta,Q^2,x_{\Pomeron})
\nonumber\\
&\times
\int d^2 b  \int^{\infty}_{-\infty}d z_1 \int^{\infty}_{z_1}d z_2 \rho_A(\vec{b},z_1) \rho_A(\vec{b},z_2) e^{i (z_1-z_2) x_{\Pomeron} m_N}
 \frac{\left\langle \sigma^2 e^{-\frac{A}{2} (1-i\eta) \sigma \int_{z_1}^{z_2} dz^{\prime} \rho_A(\vec{b},z^{\prime})} \right\rangle_j}{\langle \sigma^2 \rangle_j}
\,.
\label{eq:fluct2_c}
\end{align}

The evaluation of nuclear shadowing in the leading twist theory of nuclear shadowing, and
Eq.~(\ref{eq:fluct2_c}) in particular, does not take into account the possible 
ultra high-energy branching of the diffractive exchange which would couple to different nucleons
of the target (the so-called triple Pomeron fan or enhanced Reggeon diagrams)~\cite{Schwimmer:1975bv}.
Using the model for the interaction with $N \geq 3$ nucleons
which takes into account such diagrams,  
nuclear shadowing in nuclear PDFs was predicted in Refs.~\cite{Armesto:2003fi,Tywoniuk:2007xy}. 
  
The general form of the distribution $P_j(\sigma)$ that enters 
Eq.~(\ref{eq:fluct2_c}) is not known.
However, one can still infer the properties of
$P_j(\sigma)$ that determine the strength of nuclear shadowing.
For the virtual photon, $P_j(\sigma)$ is very broad and includes 
the states $|\sigma \rangle$ that correspond to both small and large cross sections 
$\sigma$~\cite{Frankfurt:1996ri,Frankfurt:1997zk}. The fluctuations with small cross sections 
constitute the perturbative contribution to the
photon-nucleon cross section; the fluctuations with large cross sections
correspond to the hadronic component of the virtual photon.
In practice,
$\langle \sigma^2 \rangle$ is dominated by the hadronic-size configurations.
This expectation is 
based on the QCD aligned jet model~\cite{Abramowicz:1995hb}, and
agrees well with the final analyses of the HERA data on hard diffraction which find that $\alpha_{\Pomeron} (t=0) =1.111 \pm 0.007 $~\cite{Aktas:2006hy,Aktas:2006hx}, which is practically the same as in soft processes, 
$\alpha_{\Pomeron}^{\rm soft}(0) =1.0808$~\cite{Donnachie:1992ny}. 
 Hence,
the diffractive state $X$ in Figs.~\ref{fig:Master1} and  \ref{fig:Master1b_2009}
is dominated by the large-$\sigma$ hadron-like fluctuations.
 
The key feature of 
Eq.~(\ref{eq:fluct2_c})
is that it separates the contributions of the small and large cross sections
[this was the main purpose of rewriting Eq.~(\ref{eq:fluct1}) in the form of 
Eq.~(\ref{eq:fluct2}) 
which led to Eq.~(\ref{eq:fluct2_c})]. 
While the fluctuations with large cross sections contribute 
to all moments $\langle \sigma^k \rangle$,
the fluctuations with small cross sections contribute
significantly only to 
$\langle \sigma \rangle$ and $\langle \sigma^2 \rangle$, i.e.,
to the $A xf_{j/N}(x,Q^2)$ term and the double scattering term proportional to 
$f_j^{D(3)}$.
Therefore, since the 
$\langle \dots \rangle_j/\langle \sigma^2 \rangle_j$ 
term
in Eq.~(\ref{eq:fluct2_c})
 probes the higher moments of $P_j(\sigma)$,
$\langle \sigma^k \rangle/\langle \sigma^2 \rangle$ with $k \geq 3$, it
can be evaluated with the distribution $P_j(\sigma)$, which neglects the small-$\sigma$ 
perturbative contribution and uses only the information on cross section fluctuations
from soft hadron-hadron scattering.
In particular, we assume that the relevant $P_j(\sigma)$ is equal to the distribution
over cross section fluctuations for the pion.

\section{The color fluctuation approximation}
\label{sec:cfa}

The $\langle \dots \rangle_j/\langle \sigma^2 \rangle_j$ term in 
Eq.~(\ref{eq:fluct2_c}) 
can be identically expanded in terms of $\langle \sigma^{k} \rangle_j/\langle \sigma^{2} \rangle_j$ with
$k \geq 3$. We have just explained that the required distribution $P_j(\sigma)$ is dominated
by soft hadron-like fluctuations.  For such fluctuations, 
the dispersion of $P_j(\sigma)$ 
does not lead to significant modifications of higher moments of $P_j(\sigma)$,
and it is a good approximation to use 
$\langle \sigma^{k} \rangle_j/\langle \sigma^{2} \rangle_j \approx(\langle \sigma^{3} \rangle_j/\langle \sigma^{2} \rangle_j)^{k-2}$ for all $k\geq 3$, which we shall call the 
{\it color fluctuation approximation}.
 Therefore, the $\langle \dots \rangle_j/\langle \sigma^2 \rangle_j$ term in 
Eq.~(\ref{eq:fluct2_c}) 
is expressed in terms of a single cross section, $\sigma_3^j(x,Q^2)$,
\begin{equation}
\sigma_3^j(x,Q^2) \equiv \langle \sigma^{3} \rangle_j/\langle \sigma^{2} \rangle_j=
\left(\langle \sigma^{k} \rangle_j/\langle \sigma^{2} \rangle_j \right)^{1/(k-2)}\,.
\label{eq:s_soft}
\end{equation}

Applying the color fluctuation approximation 
to Eq.~(\ref{eq:fluct2_c}), 
we obtain our final
 expression for the nuclear parton distributions modified by nuclear shadowing,
\begin{align}
&xf_{j/A}(x,Q^2)=Axf_{j/N}(x,Q^2) \nonumber\\
&-xf_{j/N}(x,Q^2) 8 \pi A (A-1)\, \Re e \frac{(1-i \eta)^2}{1+\eta^2} B_{\rm diff}
\int^{0.1}_{x} d x_{\Pomeron} \beta f_j^{D(3)}(\beta,Q^2,x_{\Pomeron})
 \nonumber\\
&\times \int d^2 b \int^{\infty}_{-\infty}d z_1 \int^{\infty}_{z_1}d z_2 \rho_A(\vec{b},z_1) \rho_A(\vec{b},z_2) e^{i (z_1-z_2) x_{\Pomeron} m_N}
 e^{-\frac{A}{2} (1-i\eta) \sigma_3^j(x,Q^2) \int_{z_1}^{z_2} dz^{\prime} \rho_A(\vec{b},z^{\prime})} \,.
\label{eq:fluct5}
\end{align}

In the 
treatment of multiple rescatterings in the leading twist theory of 
nuclear shadowing in Ref.~\cite{Frankfurt:2003zd}, 
one used the so-called quasi-eikonal approximation, which prescribes the use of 
$\sigma_3^j(x,Q^2)=\sigma_{2}^j(x,Q^2)$
in Eq.~(\ref{eq:fluct5}). 
While the quasi-eikonal and color fluctuation 
approximations
give identical results for the interaction with 
two nucleons of the nuclear target,
the color fluctuation approximation provides a more accurate treatment of the interaction
with three and more nucleons. In particular, the interaction with three nucleons is
treated exactly in
the formalism of cross section (color) fluctuations.
As follows from the definition and modeling (see below) 
of the effective rescattering cross section
$\sigma_3^j(x,Q^2)$, the color fluctuation approximation represents the scenario 
corresponding to the lower limit on nuclear shadowing within the
framework of the leading twist nuclear shadowing.

To model the distribution $P_j(\sigma)$, we assume that $P_j(\sigma)=P_{\pi}(\sigma)$, 
where $P_{\pi}(\sigma)$ is the distribution over cross sections for the pion.
This assumption  
attempts to capture the observation which we explained above that
for the higher rescattering contributions to nuclear shadowing, only large-size fluctuations
of the virtual photon wave function are important. This is a somewhat extreme assumption
which results in
the smallest nuclear shadowing 
(see Fig.~\ref{fig:LT_predictions} below).

\begin{figure}[t]
\begin{center}
\epsfig{file=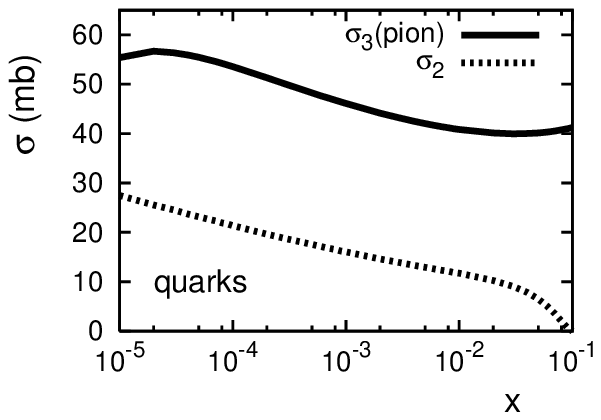,scale=1.25}
\epsfig{file=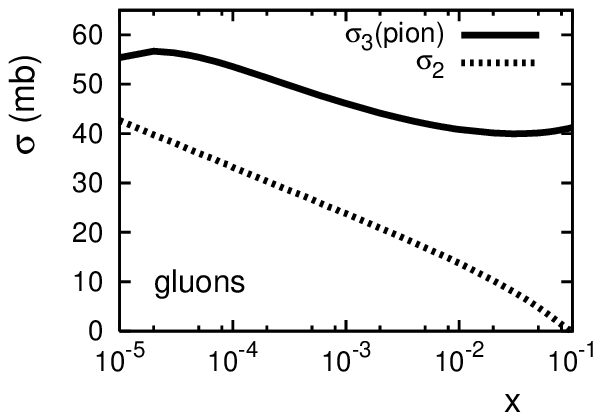,scale=1.25}
\caption{The cross sections $\sigma_3^{\rm pion}(W^2)\equiv \sigma_{3}^j(x,Q_0^2)$ [Eq.~(\ref{eq:s_soft})]
and $\sigma_{2}^j(x,Q_0^2)$ [Eq.~(\ref{eq:m17})]
as functions of Bjorken $x$ at 
$Q_0^2=4$ GeV$^2$.
The left panel corresponds to the ${\bar u}$-quark; the right panel corresponds to gluons.
}
\label{fig:sigma3_2009}
\end{center}
\end{figure}

The distribution $P_{\pi}(\sigma)$ is conveniently parameterized in the following form~\cite{Blaettel:1993rd}:
\begin{equation}
P_j(\sigma)=P_{\pi}(\sigma)=N e^{-\frac{(\sigma -\sigma_0)^2}{(\Omega \sigma_0)^2}} \,.
\label{eq:Ppion}
\end{equation}
The parameters $N$, $\sigma_0$ and $\Omega$ are constrained by following requirements:
\begin{eqnarray}
\int_0^{\infty} d \sigma P_{\pi}(\sigma) &=& 1 \,,
\nonumber\\
\int_0^{\infty} d \sigma P_{\pi}(\sigma) \sigma &=& \sigma_{\rm tot}^{\pi N}(W^2) \,,
\nonumber\\
\int_0^{\infty} d \sigma P_{\pi}(\sigma) \sigma^2 &=& \left(\sigma_{\rm tot}^{\pi N}(W^2)\right)^2 \left(1+\omega_{\sigma}(W^2)\right) \,,
\label{eq:Psigma_moments}
\end{eqnarray}
where $\sigma_{\rm tot}^{\pi N}$ is the total pion-nucleon cross section;
$\omega_{\sigma}$ is the parameter characterizing the dispersion of the distribution
$P_{\pi}(\sigma)$. Both $\sigma_{\rm tot}^{\pi N}$ and $\omega_{\sigma}$ depend on
$W^2=Q^2/x-Q^2+m_N^2$.
 In our numerical analysis, we 
used the Donnachie-Landshoff parameterization for 
$\sigma_{\rm tot}^{\pi N}$~\cite{Donnachie:1992ny}:
\begin{equation}
\sigma_{\rm tot}^{\pi N}(W^2)=\frac{1}{2}\left(\sigma_{\rm tot}^{\pi^+ N}
+\sigma_{\rm tot}^{\pi^- N}
\right)=13.63 \,(W^2)^{0.0808}+31.79\,(W^2)^{-0.4525} \ {\rm mb} \,.
\label{eq:sigma_pion}
\end{equation}
Note that in our calculations, we effectively use only the first term in Eq.~(\ref{eq:sigma_pion}),
see also Fig.~\ref{fig:sigma3_2009}.
The parameter $\omega_{\sigma}$ decreases with increasing energy~\cite{Guzey:2005tk}, 
which means that cross section fluctuations decrease with increasing energy. 
For the pion projectile, $\omega_{\sigma} \approx 0.4$ at the pion energy of
$E_{\pi} \approx 300$ GeV~\cite{Blaettel:1993ah,Blaettel:1993rd}, which corresponds to
$W^2 \approx 600$ GeV$^2$. 
At the CDF energy of $W^2=(546)^2 \approx 3 \times 10^{5}$ GeV$^2$, 
$\omega_{\sigma} \approx 0.16 \times (3/2)=0.24$, where the factor $0.16$ is $\omega_{\sigma}$
for the proton at the CDF energy~\cite{Blaettel:1993ah}
and the factor $3/2$ reflects the constituent quark counting~\cite{Blaettel:1993rd}.
Assuming a simple linear interpolation between the two energies, we arrive at the
following model for $\omega_{\sigma}$:
\begin{equation}
\omega_{\sigma}(W^2)=0.4 -0.16\, \frac{W^2-W_1^2}{W_2^2-W_1^2} \,,
\label{eq:omega_sigma}
\end{equation}
where $W_1^2=600$ GeV$^2$ and $W_2^2=3 \times 10^{5}$ GeV$^2$.
Equations~(\ref{eq:Psigma_moments}), (\ref{eq:sigma_pion}) and (\ref{eq:omega_sigma})
fully determine $P_{\pi}(\sigma)$ 
and $\sigma_{3}^j(x,Q^2)$ and their energy (Bjorken $x$) dependence.
In order to keep track of our modeling $P_j(\sigma)=P_{\pi}(\sigma)$, it is 
also convenient to introduce the notation $\sigma_3^{\rm pion}(W^2) \equiv \sigma_{3}^j(x,Q^2)$.

Figure~\ref{fig:sigma3_2009} presents 
$\sigma_3^{\rm pion}(W^2) \equiv \sigma_{3}^j(x,Q^2)$
and 
$\sigma_{2}^j(x,Q^2)$ as functions of Bjorken $x$ at fixed $Q_0^2=4$ GeV$^2$ 
($W^2=Q_0^2/x-Q_0^2+m_N^2$). 
The left panel corresponds to the ${\bar u}$-quark; the right panel corresponds to gluons.
(Note that $\sigma_{3}^j(x,Q^2)$ is flavor-independent in our model.)
The fact that $\sigma_3^{\rm pion}(W^2) > \sigma_{2}^j(x,Q_0^2)$, 
which is equivalent to $\langle \sigma^3 \rangle_j/\langle \sigma^2 \rangle_j >
\langle \sigma^2 \rangle_j/\langle \sigma \rangle_j$,
is a general property of the distribution $P_j(\sigma)$.

\section{Predictions for nuclear PDFs}
\label{sec:predictions}

Equation~(\ref{eq:fluct5}) allows one to calculate nuclear PDFs modified by nuclear shadowing.
The key 
inputs
 for this calculation are the diffractive parton distributions
$f_j^{D(3)}$ and the slope of the diffractive $\gamma^{\ast}p$ cross section $B_{\rm diff}$.
The current experimental uncertainties 
of $f_j^{D(3)}$ and $B_{\rm diff}$
lead to an uncertainty in the predictions of nuclear shadowing which
is much larger than the uncertainty associated with the use
of the color fluctuation approximation instead of  
the complete treatment of color fluctuations.
It is also important to note that $\sigma_{3}^j(x,Q^2)$ depends weakly on energy (Bjorken $x$).
Hence, measuring nuclear shadowing with one nucleus at e.g., $x=10^{-3}$, will 
further improve our predictions for all $x$ and $A$.

\begin{figure}[t]
\begin{center}
\epsfig{file=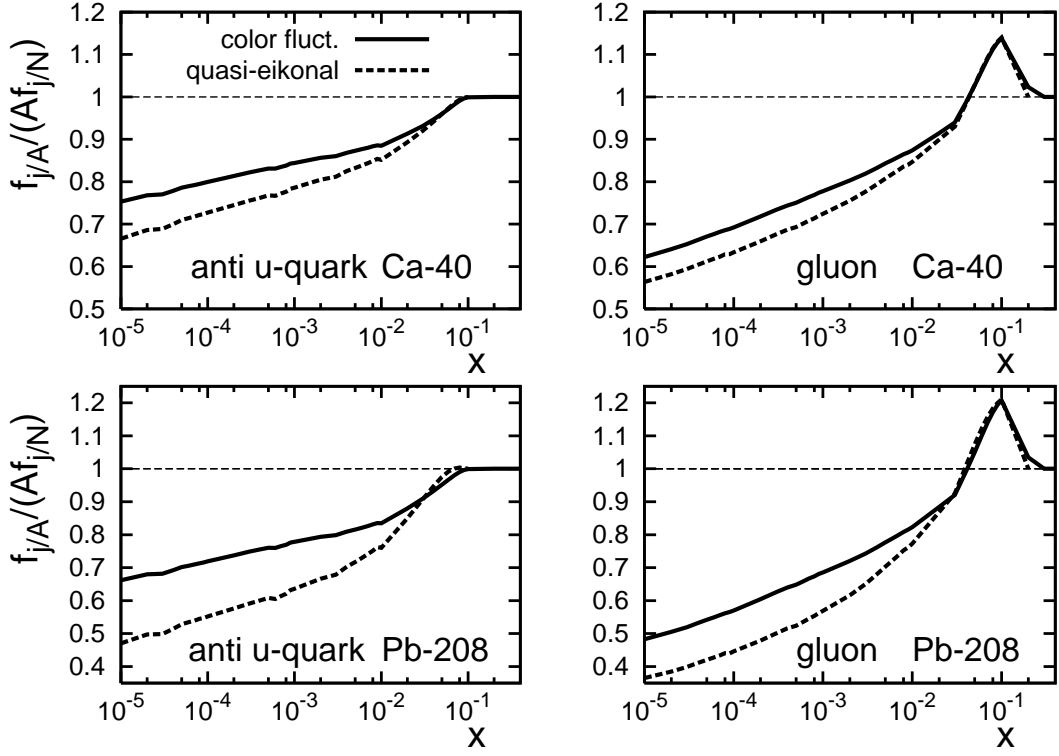,scale=1.4}
\caption{The ratio of the nuclear to nucleon parton distributions,
$f_{j/A}(x,Q^2)/[A f_{j/N}(x,Q^2)]$, as a function of Bjorken $x$ at 
the input scale $Q_0^2=4$ GeV$^2$.
The solid (dotted) curves correspond to the color fluctuation (quasi-eikonal) approximation.
The left (right) panels correspond to ${\bar u}$-quark (gluon) distributions.}
\label{fig:LT_predictions}
\end{center}
\end{figure}

Figure~\ref{fig:LT_predictions} presents the ratio of the nuclear to nucleon parton distributions,
$f_{j/A}(x,Q^2)/[A f_{j/N}(x,Q^2)]$, as a function of Bjorken $x$
at the input scale $Q_0^2=4$ GeV$^2$.
The solid curves correspond to the color fluctuation approximation, Eq.~(\ref{eq:fluct5}),
%vg
and $\sigma_{3}^j(x,Q^2) \equiv \sigma_3^{\rm pion}(W^2)$, Eqs.~(\ref{eq:Psigma_moments}), (\ref{eq:sigma_pion}) and (\ref{eq:omega_sigma}).
The dotted curves are obtained
using the quasi-eikonal approximation with $\sigma_3^{j}(x,Q^2)=\sigma_{2}^j(x,Q^2)$
in Eq.~(\ref{eq:fluct5}).
The two left panels 
correspond to ${\bar u}$-quarks; the two right panels correspond to gluons. 
Note that we added the effect of antishadowing for the gluon distribution for $0.03 \leq x \leq 0.2$ using the method described in Ref.~\cite{Frankfurt:2003zd}.  The panels in the top row are for $^{40}$Ca; the bottom 
panels are for $^{208}$Pb.

As one can see from Fig.~\ref{fig:LT_predictions}, the color fluctuation approximation 
corresponds to the smaller nuclear shadowing at the input scale $Q_0^2=4$ GeV$^2$ 
than the quasi-eikonal approximation. For instance, at $x=10^{-4}$ and
for $^{40}$Ca, ${\bar u}_{A}(x,Q_0^2)_{|{\rm cf}}/{\bar u}_A(x,Q_0^2)_{|\rm qe}=1.10$
and $g_{A}(x,Q_0^2)_{|\rm cf}/g_A(x,Q_0^2)_{|\rm qe}=1.11$ (the subscripts indicate
the color fluctuation and quasi-eikonal approximations, respectively).
At $x=10^{-4}$ and
for $^{208}$Pb, ${\bar u}_{A}(x,Q_0^2)_{|{\rm cf}}/{\bar u}_A(x,Q_0^2)_{|\rm qe}=1.31$
and $g_{A}(x,Q_0^2)_{|\rm cf}/g_A(x,Q_0^2)_{|\rm qe}=1.29$.

We described the modifications of the nuclear PDFs at small $x$ at the input scale $Q_0^2=4$ GeV$^2$.
Predictions for nuclear PDFs at higher scales $Q^2 > Q_0^2$ are obtained using the standard DGLAP 
evolution equations with the input given by Eq.~(\ref{eq:fluct5}).
As one increases $Q^2$, at a given $x$, the difference between the predictions of the 
color fluctuation and
quasi-eikonal approximations reduces: In the DGLAP evolution, the PDFs at small $x$ and large $Q^2$
are obtained from the region of larger $x$ and smaller $Q^2$, where nuclear shadowing is smaller.

One should also keep in mind that while we present our predictions for nuclear PDFs all the
way down to $x=10^{-5}$, our results for $x \lesssim 10^{-4}$ should be considered only as guiding ones
since, for this kinematic region, various effects beyond the leading twist DGLAP equation
should start becoming important.   

\section{Summary}
\label{sec:summary}

The leading twist theory of nuclear shadowing allows one to predict nuclear parton distributions
modified by nuclear shadowing. Nuclear shadowing arises as the effect 
of the interaction of a hard probe with a parton
simultaneously 
belonging to several  nucleons of the nuclear target.
While the leading twist
theory of nuclear shadowing gives unambiguous predictions for nuclear shadowing in the case when 
the interaction with only two nucleons is important, the interaction with $N \geq 3$
nucleons requires a more detailed knowledge of the dynamics of the hard diffractive processes.
 We propose a new approach to the treatment of such multiple interactions
using the formalism of cross section (color) fluctuations
 which allows us to take into account the presence of 
both point-like and average hadronic-size
configurations in the virtual photon wave function.
For practical applications, we propose a new approximation---the
color fluctuation approximation---which approximates well 
the complete treatment of color fluctuations. 
This approximation 
improves the treatment of the multiple interactions 
in the quasi-eikonal approximation, which works well in soft processes, where
color fluctuations are smaller and have a different structure 
than in DIS.
Using the developed framework, we present updated
predictions
for the effect of nuclear shadowing in nuclear parton distributions of heavy nuclei at small $x$.

\acknowledgments

The authors would like to thank L.~Frankfurt for the collaboration on the subjects presented in this work. 

Authored by Jefferson Science Associates, LLC under U.S. DOE Contract No. DE-AC05-06OR23177. The U.S. Government retains a non-exclusive, paid-up, irrevocable, world-wide license to publish or reproduce this manuscript for U.S. Government purposes.
Supported by DOE grant under contract DE-FG02-93ER40771.

\end{document}